\documentclass{cjpsuppl}% for LaTeX 2e
\begin{document}
\title{Isospectral Hamiltonians from Moyal products}
\authori{C. Figueira de Morisson Faria and A. Fring}
\addressi{Centre for Mathematical Science, City University,\\
Northampton Square, London EC1V 0HB, UK}
\authorii{}    \addressii{}
\authoriii{}   \addressiii{}
\authoriv{}    \addressiv{}
\authorv{}     \addressv{}
\authorvi{}    \addressvi{}
\headtitle{Isospectral Hamiltonians from Moyal products}
\headauthor{C. Figueira de Morisson Faria and A. Fring}
\lastevenhead{C. Figueira de Morisson Faria and A. Fring}
\pacs{03.65.-w,02.30.Mv}
\keywords{pseudo-Hermiticity, PT invariance, Moyal products}
%%%%%%%%%%%%%% Pro editory supplementu: %%%%%%%%%%%%%%%
\refnum{}%slouzi editorum pro evidenci; nakonec {}
\daterec{11 July 2006;\\final version 21 July 2006}
\suppl{A}  \year{2006} \setcounter{page}{1}
%\firstpage{1}
%\lastpage{000}
%\makefirsttitle
%%%%%%%%%%%%%%%%%%%%%%%%%%%%%%%%%%%%%%%%%%%%%%
\maketitle

\begin{abstract}
Recently Scholtz and Geyer proposed a very efficient method to compute
metric operators for non-Hermitian Hamiltonians from Moyal products. We
develop these ideas further and suggest to use a more symmetrical definition
for the Moyal products, because they lead to simpler differential equations.
In addition, we demonstrate how to use this approach to determine the
Hermitian counterpart for a Pseudo-Hermitian Hamiltonian. We illustrate our
suggestions with the explicitly solvable example of the $-x^4$-potential and
the ubiquitous harmonic oscillator in a complex cubic potential.
\end{abstract}

\section{Introduction}

Many non-Hermitian Hamiltonians $H$ are known to possess real discrete
spectra, e.g. \cite{Bender:1998ke,Znojil}, which make them potential
candidates for physical systems. Unlike as for Hermitian Hamiltonians $h$,
the conventional inner products of the corresponding wavefunctions are
usually indefinite and the central problem in this context is to construct
meaningful inner products serving to formulate a consistent quantum theory.
Besides from compensating wrong signs by hand \cite{BQZ}, there are
essentially three different, albeit in many cases equivalent, ways to
achieve this: i) by employing bi-orthonormal eigenstates \cite{Weigertbi},
ii) by constructing the so-called C-operator \cite%
{Bender:2002vv,Bender:2004sa} or iii) by restricting to pseudo-Hermitian
Hamiltonians $H$ and constructing their Hermitian counterparts $h$ related
to $H$ by a similarity transformation $\eta =\eta ^{\dagger }$ \cite%
{Most,Weigert} 
\begin{equation}
h=\eta H\eta ^{-1}=h^{\dagger }\qquad \Leftrightarrow \qquad H^{\dagger
}=\eta ^{2}H\eta ^{-2}.  \label{sim}
\end{equation}
See also \cite{Urubu} for an earlier discussion of these issues. In many
respects the last possibility is the most direct and straightforward
approach on which we will almost exclusively concentrate here. The natural
starting point is usually a given non-Hermitian Hamiltonian $H$ for which
one needs to construct $\eta $. Subsequently one may formulate all relevant
physical questions for the non-Hermitian system in terms of the conventional
formulation of quantum mechanics associated to the Hermitian system. Thus a
key task in this approach is to find $\eta ^{2}$ and $\eta $. Unfortunately
this is only possible in some exceptional cases in an exact manner and
otherwise one has to resort to perturbation theory. Building on earlier work 
\cite{Bender:2004sa,Mostafazadeh:2004qh}, we found in \cite{CA} the closed
expressions 
\begin{equation}
h=h_{0}+\sum\limits_{n=1}^{[\frac{\ell }{2}]}\frac{g^{2n}(-1)^{n}E_{n}}{%
4^{n}(2n)!}c_{q}^{(2n)}(h_{0}),\quad H=h_{0}-\sum\limits_{n=1}^{[\frac{\ell
+1}{2}]}\frac{g^{2n-1}\kappa _{2n-1}}{(2n-1)!}c_{q}^{(2n-1)}(h_{0}),
\label{HHH}
\end{equation}
which are related according to (\ref{sim}) perturbatively. Here $\left[ x%
\right] $ denotes the integer part of a number $x$, the non-Hermitian
Hamiltonian is assumed to be of the form 
\begin{equation}
H=h_{0}+igh_{1},  \label{H}
\end{equation}
$h_{0}=h_{0}^{\dagger }$, $h_{1}=h_{1}^{\dagger }$ and $c_{q}^{(n)}(x)$
denotes the $n$-fold commutator of the operator $q$ with some operator $x$.
In case one has the condition $c_{q}^{(\ell +1)}(h_{0})=0$ for some finite
integer $\ell $ the expressions (\ref{HHH}) are exact and otherwise they are
just correct up to the stated order in the coupling constant $g$. The $E_{n}$
are Euler's numbers and the $\kappa _{2n-1}$ may be constructed from them 
\cite{CA}. One should note, however, that, in practice, the above-stated
procedure may lead to rather cumbersome relations involving commutators.
This fact poses a major problem for determining closed formulae for specific
isospectral pairs of Hamiltonians, or for carrying out perturbative
computations up to higher orders.

The main purpose of this paper is to propose a practical scheme for
overcoming this difficulty. We elaborate further on a recent proposal by
Scholtz and Geyer \cite{Moyal1} to solve (\ref{sim}) by means of
Moyal products instead of computing commutators. The central idea is to
exploit isomorphic relations between commutator relations and real valued
functions multiplied by Moyal products, which correspond to differential
equations. We shall demonstrate that this approach is rather practical and
allows to compute pairs of isospectral Hamiltonians $h=h^{\dagger }$ and $%
H\neq H^{\dagger }$.

\section{Similarity transformations from Moyal Products}

\subsection{Generalities on Moyal products}

Moyal products are applied in a wide field of research, such as
non-commutative(nc) M-theory \cite{Fairlie:1997vj}, nc-string theory \cite%
{Seiberg:1999vs}, nc-integrable field theories \cite%
{Dimakis:1999nq,Grisaru:2003gt,Cabrera-Carnero:2002wq,Lechtenfeld:2004qh},
etc. The key idea is to transfer the noncommutative nature of some operators
to real valued functions. Technically one may set up such an isomorphism in
various different ways. In the present context of studying non-Hermitian
Hamiltonians such possibilities have been exploited in \cite{Moyal1}.
The authors defined the Moyal product of real valued functions depending on
the variables $x$ and $p$ as \ \ 
\begin{equation}
f(x,p)\ast g(x,p)=f(x,p)e^{i\overleftarrow{{\partial }}_{\!\!x}%
\overrightarrow{{\partial }}_{\!\!p}}g(x,p)=\sum\nolimits_{s=0}^{\infty }%
\frac{i^{s}}{s!}\partial _{x}^{s}f(x,p)\partial _{p}^{s}g(x,p).  \label{Mo1}
\end{equation}%
The classical, more widespread and symmetrical definition is, see e.g. \cite%
{Moyal,Fairlie:1998rf,Carroll} 
\begin{eqnarray}
f(x,p)\star g(x,p) &=&f(x,p)e^{\frac{i}{2}(\overleftarrow{{\partial }}%
_{\!\!x}\overrightarrow{{\partial }}_{\!\!p}-\overleftarrow{{\partial }}%
_{\!\!p}\overrightarrow{{\partial }}_{\!\!x})}g(x,p)  \label{Mo2} \\
&=&\sum\limits_{s=0}^{\infty }\frac{(-i/2)^{s}}{s!}\sum%
\limits_{t=0}^{s}(-1)^{t}\left( 
\begin{array}{r}
s \\ 
t%
\end{array}%
\right) \partial _{x}^{t}\partial _{p}^{s-t}f(x,p)\partial
_{x}^{s-t}\partial _{p}^{t}g(x,p),  \nonumber
\end{eqnarray}%
or exchange the roles of $x$ and $p$ in (\ref{Mo1}). In order to achieve a
proper isomorphism between the operator expressions and those computed with
Moyal products one requires different types of prescriptions to translate
real valued functions into operator valued expressions. Products computed
with definition (\ref{Mo1}) must be viewed as ordered products in which all $%
\hat{p}$-operators are moved to the left of all $\hat{x}$-operators. In
expressions computed with definition (\ref{Mo2}) on the other hand one
should replace each monomial $p^{m}x^{n}$ or $x^{n}p^{m}$ by the totally
symmetric polynomial $S_{m,n}$ in the $m$ operators $\hat{p}$ and $n$
operators $\hat{x}$%
\begin{equation}
\hat{S}_{m,n}=\frac{m!n!}{(m+n)!}\sum\nolimits_{\pi }\hat{p}^{m}\hat{x}^{n}.
\label{Snm}
\end{equation}%
Here $\pi $ indicates the sum over the entire permutation group. The
simplest example to illustrate this is 
\begin{equation}
\lbrack \hat{x}^{2},\hat{p}^{2}]=4i\hat{p}\hat{x}-2~~\cong ~~x^{2}\ast
p^{2}-p^{2}\ast x^{2}=4ipx-2~~\cong ~~x^{2}\star p^{2}-p^{2}\star x^{2}=4ipx.
\end{equation}%
We use here the standard canonical commutation relation $[\hat{x},\hat{p}]=i$
and throughout the paper we employ atomic units $\hbar =e=m_{e}=c\alpha =1$.
We observe that the $\ast $-product yields the correct operator expression
upon replacing $x\rightarrow \hat{x}$, $p\rightarrow \hat{p}$. The $\star $%
-product on the other hand corresponds to the correct commutation relations
for $px\rightarrow S_{1,1}=(\hat{p}\hat{x}+\hat{x}\hat{p})/2$. The defining
relation (\ref{Mo2}) is slightly more complicated than (\ref{Mo1}), but it
is more symmetrical and leads therefore to cancellations of various terms as
one can easily convince oneself. Loosely speaking, it has the advantage that
it incorporates already more of the noncommutative nature of $\hat{x}$ and $%
\hat{p}$ from the very beginning. As we shall see below, this is the reason
why it leads to simpler differential equations for the quantities we wish to
determine. These statements are also supported by comparing the equations
resulting from (\ref{Mo2}) and (\ref{Mo1}) for all examples calculated up to
now with this method \cite{Moyal1}\footnote{%
For instance for the example $H=p^{2}+igx^{3}$, definition (\ref{Mo1})
yields 
\[
2gx^{3}\eta ^{2}(x,p)+3igx^{2}\partial _{p}\eta ^{2}(x,p)-3gx\partial
_{p}^{2}\eta ^{2}(x,p)-ig\partial _{p}^{3}\eta ^{2}(x,p)+2p\partial _{x}\eta
^{2}(x,p)+i\partial _{x}^{2}\eta ^{2}(x,p)=0,
\]%
whereas (\ref{Mo2}) gives the simpler form $4gx^{3}\eta ^{2}(x,p)-3gx\partial _{p}^{2}\eta ^{2}(x,p)+4p\partial _{x}\eta
^{2}(x,p)=0.$
}.

\subsection{Hermitian counterparts from Pseudo-Hermitian Hamiltonians}

Let us now briefly discuss in general how we proceed to construct $h$ from a
given $H$ by first solving (\ref{sim}) for $\eta ^{2}$ and $\eta $. The
explicit knowledge of $\eta $ is vital, since once it is known one may
control the entire quantum mechanical formalism, such as inner products,
observables, time evolution, etc. However, in general one can not compute $%
\eta $ exactly and has to rely on perturbative methods \cite%
{Bender:2004sa,Mostafazadeh:2004qh,CA} in which one has to solve the
commutator relations occurring in (\ref{HHH}) order by order. This is a very
cumbersome procedure and up-to-now it has only been carried out for few
cases to lowest order. Here we present a simple and more efficient scheme
which leads to the exact determination of $\eta $ by employing Moyal
products. We build on suggestions of \cite{Moyal1}, but as a starting
point we use instead of the definition (\ref{Mo1}) the definition (\ref{Mo2}%
) and write the second equation in (\ref{sim}) isomorphically as 
\begin{equation}
H^{\dagger }\star \eta ^{2}=\eta ^{2}\star H.  \label{He}
\end{equation}%
With definition (\ref{Mo2}) this will then lead to a differential equation
in $\eta ^{2}$, whose order depends on the degree of $x$ and $p$ in $H$. In
most cases this equation can not be solved exactly and under these
circumstances we assume a perturbative expansion for $\eta ^{2}$ in the
coupling constant $g$, which was introduced in (\ref{H}) 
\begin{equation}
\eta ^{2}(x,p)=\sum\nolimits_{n=0}^{\infty }g^{n}c_{n}(x,p).  \label{3}
\end{equation}%
Combining (\ref{He}) and (\ref{3}) then leads to a differential equation,
which involves the functions $c_{n}$ in a recursive manner and can therefore
be solved order by order. Ultimately we aim at exact expressions, which
yield $\eta ^{2}$ to all orders in perturbation theory \cite{CAprep}. Having
then solved various differential equations, we naturally expect some
ambiguities in the general solutions, which mirror the possibility of
different boundary conditions. We would like to stress that this is not a
drawback, which is only present when using Moyal products, but it is rather
a reflection of a general feature of perturbation theory. The same kind of
ambiguity occurs in the perturbative approach based on commutators. In that
context one may only fix the operator corresponding to the $c_{n}(x,p)$ up
to any operator which commutes with the Hermitian part of $H$, that is $h_{0}
$. This means that, in (\ref{HHH}), the expressions are insensitive to any
replacement $q\rightarrow q+\tilde{q}$ with $[\tilde{q},h_{0}]=0$. A further
type of ambiguity, which is always present irrespective of an exact or
perturbative treatment, is a multiplication of $\eta ^{2}$ by operators
which commute with $H$, i.e. we could re-define $\eta ^{2}\rightarrow \eta
^{2}Q$ for any $Q$, which satisfies $[Q,H]=0$.

We may fix these ambiguities by invoking a further property of $\eta $. As
discussed in \cite{CA}, assuming a dependence on the coupling constant $g$
of the form 
\begin{equation}
\eta (-g)=\eta (g)^{-1},\qquad h(g)=h(-g)\qquad \mathrm{and}\qquad
H^{\dagger }(g)=H(-g),  \label{hg}
\end{equation}%
will guarantee the pseudo-Hermiticity relations (\ref{sim}). Therefore we
require next 
\begin{equation}
\eta ^{2}(g)\star \eta ^{2}(-g)=1,  \label{et}
\end{equation}%
which may be solved systematically order by order when we already know the
expansion (\ref{3}) up to the ambiguities. Note that (\ref{et}) is
automatically satisfied if we take $\eta ^{2}=e^{q}$ and $%
q=\sum\nolimits_{n=1}^{\infty }g^{2n-1}q_{2n-1}$ as assumed in many cases on
the grounds of PT-invariance \cite{Bender:2004sa}. Having determined $\eta
^{2}$ the transformation $\eta $ is subsequently computed easily order by
order from 
\begin{equation}
\eta \star \eta =\eta ^{2}=\sum\nolimits_{n=0}^{\infty }g^{n}c_{n}(x,p),
\label{ee}
\end{equation}%
when assuming a further power series expansion 
\begin{equation}
\eta (x,p)=\sum\nolimits_{n=0}^{\infty }g^{n}q_{n}(x,p).  \label{eet}
\end{equation}%
Finally, having obtained an explicit expression for $\eta $, we may compute
the Hermitian counterpart to $H$ from (\ref{sim}) as 
\begin{equation}
h\star \eta =\eta \star H.  \label{hk}
\end{equation}%
According to the above arguments, i.e. the second equation in (\ref{hg}), we
should find an expression of the general form 
\begin{equation}
h(x,p)=\sum\nolimits_{k=0}^{\infty }g^{2k}h_{2k}(x,p).  \label{hh}
\end{equation}%
\noindent Let us now illustrate the above with two explicit examples, by
starting with one for which the differential equation may be solved exactly
and thereafter complicating it to a case which requires perturbation theory.

\section{The non-Hermitian -x$^{4}$ potential}

As a straightforward example we consider the non-Hermitian Hamiltonian 
\begin{equation}
H(\hat{x},\hat{p})=\hat{p}^{2}-\frac{\hat{p}}{2}+\alpha \left( \hat{x}%
^{2}-1\right) \,+ig\,\left( \frac{\{\hat{x},\hat{p}^{2}\}}{2}-2\,\alpha \hat{%
x}\,\right) ,  \label{aq}
\end{equation}
which results from $H=-d^{2}/dz^{2}-\varepsilon z^{4}$ when using $z=-2i%
\sqrt{1+ix}$ as transformation, $\alpha =16\varepsilon $ and the
introduction of the coupling constant $g$ to separate off the non-Hermitian
part \cite{JM}. The exact similarity transformation $\eta $ for this
Hamiltonian was recently constructed by Jones and Mateo \cite{JM} using
perturbation theory in terms of commutators in the spirit of equation (\ref%
{HHH}). See also \cite{JM} for further reasoning on how the Hamiltonian (\ref%
{aq}) can be used to make sense of the $-\varepsilon z^{4}$-potential
despite its unappealing property of being unbounded from below.

In order to illustrate the method let us see how to determine $\eta $ by
using Moyal products. First we notice that we wish to treat $H(\hat{x},\hat{p%
})$ as a real valued function and we therefore have to replace the
anti-commutator with the appropriate Moyal products. When using definition (%
\ref{Mo2}) we have to replace $\{\hat{x},\hat{p}^{2}\}$ by $x\star
p^{2}+p^{2}\star x=2xp^{2}$ and the differential equation (\ref{He}) for the
Hamiltonian (\ref{aq}) reads 
\begin{eqnarray}
0 &=&4gp^{2}x\eta ^{2}(x,p)-8gx\alpha \eta ^{2}(x,p)-4x\alpha \partial
_{p}\eta ^{2}(x,p)  \label{d1} \\
&&-\partial _{x}\eta ^{2}(x,p)+4p\partial _{x}\eta ^{2}(x,p)+2gp\partial
_{p}\partial _{x}\eta ^{2}(x,p)-gx\partial _{x}^{2}\eta ^{2}(x,p).~~ 
\nonumber
\end{eqnarray}
As a comparison we also present the differential equation resulting from (%
\ref{He}) when using the $\ast $-product instead of the $\star $-product.
When converting the operator valued Hamiltonians into a function, we have to
pay attention to the fact that $(f\ast g)^{\ast }\neq f^{\ast }\ast g^{\ast
} $. Thus \ in $H$ we replace $i\{\hat{x},\hat{p}^{2}\}$ by $i(x\ast
p^{2}+p^{2}\ast x)=2ip^{2}x$ $-2p$, whereas in $H^{\dagger }$ we substitute $%
-i\{\hat{x},\hat{p}\}$ by $-i(x\ast p^{2}+p^{2}\ast x)=-2ip^{2}x$ $+2p$,
which is of course not the same as converting first the anti-commutator with
a subsequent conjugation. The resulting differential equation reads 
\begin{eqnarray}
0 &=&4gp^{2}x\eta ^{2}(x,p)-8gx\alpha \eta ^{2}(x,p)-4x\alpha \partial
_{p}\eta ^{2}(x,p)  \label{d2} \\
&&+i2gp^{2}\partial _{p}\eta ^{2}(x,p)-i2\alpha \partial _{p}^{2}\eta
^{2}(x,p)-i4g\alpha \partial _{p}\eta ^{2}(x,p)+i4gp\eta ^{2}(x,p)  \nonumber
\\
&&-(1+2g-4p-4pxig)\partial _{x}\eta ^{2}(x,p)+(2i-2gx)\partial _{x}^{2}\eta
^{2}(x,p).  \nonumber
\end{eqnarray}
Obviously, equation (\ref{d2}) is more complicated than (\ref{d1}), which
illustrates our assertion that the $\star $-product is more advantageous
than the $\ast $-product. Ultimately they should lead, however, to the same
result. Indeed, each line in (\ref{d1}) as well as (\ref{d2}) vanishes
separately for 
\begin{equation}
\eta ^{2}(x,p)=e^{\frac{g\,p^{3}}{3\,\alpha }-2\,g\,p},
\end{equation}
which, when compensating a slight difference in convention, is precisely the
same expression as found in \cite{JM}. To find $\eta (x,p)$ from $\eta
^{2}(x,p)$ by means of (\ref{ee}) is trivial in this case as we just have to
take the square root, i.e. $\eta (x,p)=e^{\frac{g\,p^{3}}{6\,\alpha }-g\,p}$%
. Using (\ref{hk}) thereafter we find the Hermitian counterpart of $H$%
\begin{equation}
h(x,p,g)=p^{2}-\frac{p}{2}+\alpha \left( x^{2}-1\right) \,+g^{2}\frac{%
\,\left( p^{2}-2\,\alpha \right) ^{2}}{4\,\alpha }.
\end{equation}
When setting the artificially introduced parameter $g$ to $1$, we recover
precisely the expression found in \cite{JM} 
\begin{equation}
h(x,p,g=1)=\frac{p^{4}}{4\,\alpha }-\frac{p}{2}+x^{2}\,\alpha .
\end{equation}
In \cite{JM} also the interesting massive case $H=-d^{2}/dz^{2}+m^{2}z^{2}-%
\varepsilon z^{4}$ has been discussed. The transformation from the $z$%
-variable to the $x$-variable will add in (\ref{aq}) a term $-m^{2}(1+4igx)$%
. The resulting similarity transformation is then $\eta (x,p)=e^{\frac{%
g\,p^{3}}{6\,\alpha }-g\,p(1+2m^{2}/\alpha )}$ and the Hermitian counterpart
reads 
\begin{equation}
h(x,p,g)=p^{2}-\frac{p}{2}+\alpha \left( x^{2}-1\right) \,-4m^{2}+g^{2}\frac{%
\,\left( p^{2}-2\,\alpha -4\,m^{2}\right) ^{2}}{4\,\alpha },
\end{equation}
which for $g\rightarrow 1$ reduces precisely to the expression reported in 
\cite{JM}.

Let us next embark on an example for which the differential equation (\ref%
{d1}) can not be solved exactly.

\section{Harmonic oscillator perturbed by a complex cubic potential}

The prototype example for the study of non-Hermitian Hamiltonian systems is
the harmonic oscillator perturbed with a complex cubic potential 
\begin{equation}
H=\frac{p^{2}}{2}+\frac{x^{2}}{2}+igx^{3}.  \label{qcube}
\end{equation}%
It was the discovery \cite{Bender:1998ke} that this Hamiltonian possesses a
positive real discrete spectrum, which led to the current interest in this
subject. This Hamiltonian is obviously non-Hermitian, but PT-invariant and
pseudo-Hermitian \cite{Bender:2002vv,Weigert,Most,Mostafazadeh:2004qh}. The
latter property means that the relations (\ref{sim}) hold. For the
Hamiltonian (\ref{qcube}) the conjugation relation (\ref{He}) translates
into the differential equation\footnote{%
For comparison definition (\ref{Mo1}) yields 
\begin{eqnarray*}
0 &=&4\,g\,x^{3}\,\eta ^{2}(x,p)-2\,x\,\partial _{p}\eta
^{2}(x,p)+6\,i\,g\,x^{2}\,\partial _{p}\eta ^{2}(x,p)-i\partial
_{p}^{2}\,\eta ^{2}(x,p) \\
&&-6\,g\,x\,\partial _{p}^{2}\eta ^{2}(x,p)-2\,i\,g\,\partial _{p}^{3}\eta
^{2}(x,p)+2\,p\,\partial _{x}\eta ^{2}(x,p)+i\,\partial _{x}^{2}\eta
^{2}(x,p).
\end{eqnarray*}%
} 
\begin{equation}
0=4gx^{3}\eta ^{2}(x,p)-3gx\partial _{p}^{2}\eta ^{2}(x,p)-2x\partial
_{p}\eta ^{2}(x,p)+2p\partial _{x}\eta ^{2}(x,p).
\end{equation}%
Using the expansion (\ref{ee}) we find the recursive equation 
\begin{equation}
0=4x^{3}c_{n-1}(x,p)-3x\partial _{p}^{2}c_{n-1}(x,p)-2x\partial
_{p}c_{n}(x,p)+2p\partial _{x}c_{n}(x,p).  \label{ccc}
\end{equation}%
Notice the occurrence of the aforementioned ambiguities. It is easy to see
that we may add to $c_{n}(x,p)$ any arbitrary function $\zeta \lbrack
(p^{2}+x^{2})/2]$. This type of ambiguity just reflects the fact that in the
perturbative formulation in terms of operators we may add to $q$ any
function of $h_{0}=(p^{2}+x^{2})/2$. We write $c_{n}(x,p)=\tilde{c}%
_{n}(x,p)+\zeta \lbrack (p^{2}+x^{2})/2]$ and determine first the functions $%
\tilde{c}_{n}(x,p)$ which also obey (\ref{ccc}).

Taking as the initial condition $\tilde{c}_{0}(x,p)=1$, we find order by
order

\begin{eqnarray*}
\tilde{c}_{1}(x,p) &=&\frac{4\,p^{3}}{3}+2\,p\,x^{2},\qquad \tilde{c}%
_{2}(x,p)=6\,x^{2}-\frac{2\,p^{2}\,x^{4}}{3}-\frac{8\,x^{6}}{9}, \\
\tilde{c}_{3}(x,p) &=&\!\!\frac{112p^{5}}{15}-\frac{64p^{9}}{81}+\frac{%
56p^{3}x^{2}}{3}-\frac{32p^{7}x^{2}}{9}\!+14px^{4}-\frac{56p^{5}x^{4}}{9}\!-%
\frac{140p^{3}x^{6}}{27} \\
&&-\frac{16px^{8}}{9}, \\
\tilde{c}_{4}(x,p) &=&112p^{2}x^{2}-\frac{128p^{6}x^{2}}{3}+98x^{4}-\frac{%
1856p^{4}x^{4}}{15}+\frac{32p^{8}x^{4}}{81}-\frac{5392p^{2}x^{6}}{45} \\
&&+\frac{416p^{6}x^{6}}{243}-\frac{1768x^{8}}{45}+\frac{230p^{4}x^{8}}{81}+%
\frac{176p^{2}x^{10}}{81}+\frac{160x^{12}}{243}, \\
\tilde{c}_{5}(x,p) &=&-224p^{3}+\frac{2752p^{7}}{5}-\frac{19456p^{11}}{405}+%
\frac{2048p^{15}}{3645}-336px^{2}+\frac{9632p^{5}x^{2}}{5} \\
&&-\frac{107008p^{9}x^{2}}{405}+\frac{1024p^{13}x^{2}}{243}+1928p^{3}x^{4}-%
\frac{26752p^{7}x^{4}}{45}+\frac{3328p^{11}x^{4}}{243} \\
&&+\frac{8332px^{6}}{15}-\frac{272896p^{5}x^{6}}{405}+\frac{18304p^{9}x^{6}}{%
729}-\frac{154132p^{3}x^{8}}{405}+\frac{760p^{7}x^{8}}{27} \\
&&-\frac{11488px^{10}}{135}+\frac{23524p^{5}x^{10}}{1215}+\frac{%
5536p^{3}x^{12}}{729}+\frac{320px^{14}}{243},
\end{eqnarray*}%
This can be expanded effortlessly to higher orders, but we will not report
these functions here. Next we demand the dependence on the coupling constant 
$g$ to be of the form (\ref{et}) in order to fix the ambiguities. Since we
may add to $c_{a}(x,p)$ any arbitrary function $\zeta
_{0}^{a}[h_{0}=(p^{2}+x^{2})/2]\,$, due to the recursive equation (\ref{ccc}%
) this function will produce descendents at higher level in $c_{n}$ for $n>a$%
. Thus we have the general form 
\begin{equation}
c_{n}(x,p)=\tilde{c}_{n}(x,p)+\sum\nolimits_{k=0}^{n-a}\zeta
_{k}^{(a)}[h_{0}].
\end{equation}%
We can now use this function and achieve that (\ref{et}) is satisfied order
by order. Explicitly, 
\begin{eqnarray*}
c_{1}(x,p) &=&\frac{4p^{3}}{3}+2px^{2},\quad  \\
c_{2}(x,p) &=&\tilde{c}_{2}(x,p)+\zeta _{0}^{(2)}[h_{0}]=\tilde{c}_{2}(x,p)+%
\frac{8}{9}h_{0}^{3}-4h_{0}.
\end{eqnarray*}%
As we note, indeed the additional function just depends on $h_{0}$. In a
similar fashion we can compute the higher order functions 
\begin{eqnarray*}
c_{3}(x,p) &=&12p-\frac{248p^{5}}{15}+\frac{32p^{9}}{81}-\frac{64p^{3}x^{2}}{%
3}+\frac{16p^{7}x^{2}}{9}-2px^{4}+\frac{8p^{5}x^{4}}{3}+\frac{4p^{3}x^{6}}{3}%
, \\
c_{4}(x,p) &=&152p^{4}-\frac{832p^{8}}{45}+\frac{32p^{12}}{243}+(56p^{2}-%
\frac{2368p^{6}}{45}+\frac{64p^{10}}{81})x^{2} \\
&&-(26+\frac{128p^{4}}{3}-\frac{16p^{8}}{9})x^{4}-(8p^{2}-\frac{16p^{6}}{9}%
)x^{6}+\frac{2p^{4}x^{8}}{3}, \\
c_{5}(x,p) &=&-1024p^{3}+\frac{2144p^{7}}{5}-\frac{4672p^{11}}{405}+\frac{%
128p^{15}}{3645}+(-168p+\frac{10864p^{5}}{15} \\
&&-\frac{20416p^{9}}{405}+\frac{64p^{13}}{243})x^{2}+(\frac{712p^{3}}{3}-%
\frac{3488p^{7}}{45}+\frac{64p^{11}}{81})x^{4} \\
&&-(28p+48p^{5}-\frac{32p^{9}}{27})x^{6}-(\frac{28p^{3}}{3}-\frac{8p^{7}}{9}%
)x^{8}+\frac{4p^{5}x^{10}}{15}, \\
c_{6}(x,p) &=&3584p^{2}-\frac{98336p^{6}}{15}+\frac{340256p^{10}}{675}-\frac{%
6016p^{14}}{1215}+\frac{256p^{18}}{32805}-(1024 \\
&&+6160p^{4}-\frac{70592p^{8}}{45}+\frac{35392p^{12}}{1215}-\frac{256p^{16}}{%
3645})x^{2}-(216p^{2}-\frac{70288p^{6}}{45} \\
&&+\frac{26816p^{10}}{405}-\frac{64p^{14}}{243})x^{4}+(340+\frac{1520p^{4}}{3%
}-\frac{3232p^{8}}{45}+\frac{128p^{12}}{243})x^{6} \\
&&+(14p^{2}-\frac{328p^{6}}{9}+\frac{16p^{10}}{27})x^{8}-(\frac{20p^{4}}{3}-%
\frac{16p^{8}}{45})x^{10}+\frac{4p^{6}x^{12}}{45},
\end{eqnarray*}%
We observed that always $\zeta _{0}^{(2n+1)}[h_{0}]=0$. Making next the
ansatz (\ref{eet}), we compute the $q_{n}(x,p)$ by reading off the
corresponding powers in the expression (\ref{ee}). We find order by order 
\begin{eqnarray*}
q_{1}(x,p) &=&\frac{2\,p^{3}}{3}+px^{2},\quad q_{2}(x,p)=-p^{2}+\frac{2p^{6}%
}{9}+\frac{x^{2}}{2}+\frac{2p^{4}x^{2}}{3}+\frac{p^{2}x^{4}}{2}, \\
q_{3}(x,p) &=&6p-\frac{79p^{5}}{15}+\frac{4p^{9}}{81}-\frac{23p^{3}x^{2}}{3}+%
\frac{2p^{7}x^{2}}{9}-\frac{13px^{4}}{4}+\frac{p^{5}x^{4}}{3}+\frac{%
p^{3}x^{6}}{6}, \\
q_{4}(x,p) &=&\frac{67p^{4}}{2}-\frac{148p^{8}}{45}+\frac{2p^{12}}{243}+(%
\frac{37p^{2}}{2}-\frac{442p^{6}}{45}+\frac{4p^{10}}{81})x^{2} \\
&&-(\frac{61}{8}+\frac{29p^{4}}{3}-\frac{p^{8}}{9})x^{4}-(\frac{7p^{2}}{2}-%
\frac{p^{6}}{9})x^{6}+\frac{p^{4}x^{8}}{24},\quad  \\
q_{5}(x,p) &=&\frac{-997p^{3}}{3}+\frac{443p^{7}}{5}-\frac{434p^{11}}{405}+%
\frac{4p^{15}}{3645}-\frac{355px^{2}}{2}+\frac{4681p^{5}x^{2}}{30} \\
&&-\frac{1952p^{9}x^{2}}{405}+\frac{2p^{13}x^{2}}{243}+\frac{1199p^{3}x^{4}}{%
12}-\frac{361p^{7}x^{4}}{45}+\frac{2p^{11}x^{4}}{81} \\
&&+\frac{149px^{6}}{8}-6p^{5}x^{6}+\frac{p^{9}x^{6}}{27}-\frac{43p^{3}x^{8}}{%
24}+\frac{p^{7}x^{8}}{36}+\frac{p^{5}x^{10}}{120}, \\
q_{6}(x,p) &=&\frac{1677p^{2}}{2}-\frac{103649p^{6}}{90}+\frac{84373p^{10}}{%
1350}-\frac{286p^{14}}{1215}+\frac{4p^{18}}{32805}-(\frac{1131}{4} \\
&&+\frac{16811p^{4}}{12}-\frac{8987p^{8}}{45}+\frac{1717p^{12}}{1215}-\frac{%
4p^{16}}{3645})x^{2}-(\frac{2503p^{2}}{8}-\frac{43021p^{6}}{180} \\
&&+\frac{1361p^{10}}{405}-\frac{p^{14}}{243})x^{4}+(\frac{1861}{16}+\frac{%
1625p^{4}}{12}-\frac{359p^{8}}{90}+\frac{2p^{12}}{243})x^{6} \\
&&+(\frac{943p^{2}}{32}-\frac{173p^{6}}{72}+\frac{p^{10}}{108})x^{8}-(\frac{%
29p^{4}}{48}-\frac{p^{8}}{180})x^{10}+\frac{p^{6}x^{12}}{720}
\end{eqnarray*}%
Finally, having obtained an explicit expression for the $\eta $, we may
compute the Hermitian counterpart to $H$ from (\ref{hk}). We find order by
order 
\begin{eqnarray}
h_{0}(\hat{x},\hat{p}) &=&\frac{\hat{p}^{2}}{2}+\frac{\hat{x}^{2}}{2},\quad
h_{2}(\hat{x},\hat{p})=-\frac{1}{2}+3\hat{S}_{2,2}+\frac{3\,\hat{x}^{4}}{2} 
\nonumber \\
h_{4}(\hat{x},\hat{p}) &=&27\hat{p}^{2}+2\hat{p}^{6}+\frac{15\hat{x}^{2}}{2}%
-36\hat{S}_{4,2}-\frac{51\hat{S}_{2,4}}{2}-\frac{7\,\hat{x}^{6}}{2} \\
h_{6}(\hat{x},\hat{p}) &=&128-984\hat{p}^{4}-72\hat{p}^{8}-1464\hat{S}%
_{2,2}+768\hat{S}_{6,2}-84\,\hat{x}^{4}+660\hat{S}_{4,4}  \nonumber \\
&&+288\hat{S}_{2,6}+\frac{27\hat{x}^{8}}{2},  \nonumber
\end{eqnarray}%
where we have already converted from real valued functions to operators.
Notice that $h(\hat{x},\hat{p})$ is of the form (\ref{hh}) and as we expect
all terms $h_{2k-1}=0$, such that indeed the second relation in (\ref{hg})
holds. Up to $h_{4}(\hat{x},\hat{p})$ our results agree precisely with \cite%
{HJ}. Higher orders do not seem to appear in the literature.
\section{Conclusions}
We have shown that Moyal products can be used as a very powerful tool to
compute the similarity transformation $\eta $ relating a pseudo-Hermitian
Hamiltonian $H$ to its Hermitian counterpart $h$. As concrete examples, we
have applied such products to a situation for which $\eta $ can be
calculated exactly, namely the non-Hermitian $-x^{4}$ potential, and to a
case for which it can only be computed perturbatively, that is the harmonic
oscillator with a complex cubic perturbation. In the latter case, we
profited considerably from the fact that the Moyal products involve only
real-valued functions, instead of commutators, and lead to differential
equations which can be solved recursively. Relation (\ref{et}) was crucial
to fix the ambiguities, which result from operators which commute with the
unperturbed Hamiltonian $h_{0}$.

In a more general context, this is a far less cumbersome procedure than
those involving commutators, and allows one to evaluate isospectral
pseudo-Hermitian--Hermitian pairs perturbatively up to a very high order.
Even more, Moyal products can be employed to obtain closed formula for the
similarity transformation $\eta $ to all orders in perturbation theory. Such
formulae have been obtained for several specific isospectral pairs, and will
be discussed in a subsequent publication \cite{CAprep}.

Despite the practical use of this approach, there are clear limitations,
such as for instance when the potential is of a nonpolynomial nature. In
that case the sum in the definition for the Moyal product does not terminate
and one has to deal with differential equations of infinite order. Hence,
there is a clear need for further alternative methods, such as recently
proposed in \cite{MOT}.

\textbf{Acknowledgments}. We are grateful to Hugh Jones for useful comments.

%\lastevenpage


\begin{thebibliography}{99}
\bibitem{Bender:1998ke} C.~M.~Bender and S.~Boettcher, \newblock Phys. Rev.
Lett. \textbf{80}, 5243 (1998).

\bibitem{Znojil} M.~Znojil, \newblock Phys. Lett. \textbf{A259}, 220 (1999);
Phys. Lett. \textbf{A264}, 108 (1999).

\bibitem{BQZ} B.~Bagchi, C.~Quesne and M.~Znojil, \newblock  Mod. Phys.
Lett. \textbf{A16} 2047 (2001).

\bibitem{Weigertbi} S.~Weigert, \newblock Phys. Rev. \textbf{A68}, 062111(4)
(2003).

\bibitem{Bender:2002vv} C.~M. Bender, D.~C. Brody, and H.~F. Jones, %
\newblock Phys. Rev. Lett. \textbf{89}, 270401(4) (2002).

\bibitem{Bender:2004sa} C.~M. Bender, D.~C. Brody, and H.~F. Jones, %
\newblock Phys. Rev. \textbf{D70}, 025001(19) (2004).

\bibitem{Most} A.~Mostafazadeh, \newblock J. Math. Phys. \textbf{43}, 205
(2002); J. Math. Phys. \textbf{43}, 2814 (2002); J. Math. Phys. \textbf{43},
3944 (2002); J. Phys. \textbf{A36}, 7081 (2003).

\bibitem{Weigert} S.~Weigert, \newblock J. Phys. \textbf{B5}, S416 (2003).

\bibitem{Urubu} F.~G. Scholtz, H.~B. Geyer and F.~J.~W. Hahne, \newblock %
Ann. Phys. \textbf{213}, 74 (1992).

\bibitem{Mostafazadeh:2004qh} A.~Mostafazadeh, \newblock J. Phys. \textbf{A38%
}, 6557 (2005).

\bibitem{CA} C.~Figueira~de Morisson~Faria and A.~Fring, \newblock  J. Phys. 
\textbf{A39}, 9269 (2006).

\bibitem{Moyal1} F.~G.~Scholtz and H.~B.~Geyer, \newblock Phys. Lett. 
\textbf{B634}, 84 (2006); quant-ph/0602187.

\bibitem{CAprep} C.~Figueira~de Morisson~Faria and A.~Fring, \newblock in
preparation.

\bibitem{Moyal} J.~E.~Moyal, Proc. Cambridge Phil. Soc. \textbf{45}, 99
(1949).

\bibitem{Fairlie:1997vj} D.~B. Fairlie, \newblock Mod. Phys. Lett. \textbf{%
A13}, 263 (1998).

\bibitem{Seiberg:1999vs} N.~Seiberg and E.~Witten, \newblock JHEP \textbf{09}%
, 032 (1999).

\bibitem{Dimakis:1999nq} A.~Dimakis and F.~M{\"{u}}ller-Hoissen, \newblock %
Int. J. Mod. Phys. \textbf{B14}, 2455 (2000).

\bibitem{Grisaru:2003gt} M.~T. Grisaru and S.~Penati, \newblock Nucl. Phys. 
\textbf{B655}, 250 (2003).

\bibitem{Cabrera-Carnero:2002wq} I.~Cabrera-Carnero and M.~Moriconi, %
\newblock Nucl. Phys. \textbf{B673}, 437 (2003).

\bibitem{Lechtenfeld:2004qh} O.~Lechtenfeld, L.~Mazzanti, S.~Penati, A.~D.
Popov, and L.~Tamassia, \newblock Nucl. Phys. \textbf{B705}, 477 (2005).

\bibitem{Fairlie:1998rf} D.~B. Fairlie, \newblock J. of Chaos, Solitons and
Fractals \textbf{10}, 365 (1999).

\bibitem{Carroll} R.~Carroll, \newblock North-Holland Mathematics Studies 
\textbf{186}, (Elsevier, Amsterdam) (2000).

\bibitem{JM} H.~F. Jones and J.~Mateo, \newblock  Phys. Rev. \textbf{D73},
085002(2006).

\bibitem{HJ} H.~Jones, \newblock J. Phys. \textbf{A38}, 1741 (2005).

\bibitem{MOT} A.~Mostafazadeh, \newblock quant-ph/0603023 (2006).
\end{thebibliography}
\end{document}